**Extended Depth of Field Magneto-Optical Kerr Microscopy for Applications in 3D Nanomagnetism**

Le Zhao,[1,*] Alexander Rabensteiner,[1] Miguel Ángel Cascales-Sandoval,[1] Naëmi Leo,[1,2] Sabri Koraltan,[1] and Amalio Fernández-Pacheco[1,†]

[1]Institute of Applied Physics, TU Wien, Wiedner Hauptstraße 8-10, 1040 Vienna, Austria
[2]Department of Physics, Loughborough University, Epinal Way, Loughborough LE11 3TU, United Kingdom

*Le Zhao: le.zhao@tuwien.ac.at
†Amalio Fernández-Pacheco: amalio.fernandez-pacheco@tuwien.ac.at

**ABSTRACT** High-resolution imaging of magnetic nanostructures is essential for understanding fundamental spin phenomena and designing advanced devices. Recent developments in three-dimensional (3D) nanomagnetism have highlighted the growing need for imaging techniques that can capture magnetic structures across curved or tilted surfaces with high sensitivity. Though widely used as a powerful technique for imaging magnetization states, conventional magneto-optical Kerr effect (MOKE) microscopy faces limitations in measuring non-planar systems due to its shallow depth of field (DOF). Here, we present an extended depth of field MOKE imaging approach, combining through-focus scanning with image-stitching-based reconstruction, to obtain sharp and well-resolved magnetic domain images across non-planar sample geometries. The method is validated on both perpendicularly and in-plane magnetized films tilted on purpose for this study, enabling quantitative analysis of domain morphology and detection sensitivity. This laboratory-accessible technique provides a fast and versatile route for 3D magnetic imaging, complementing large-scale X-ray-based methods and offering a practical tool for investigating non-planar magnetic samples with tilted or curved 3D geometries.

## I. INTRODUCTION

The exploration of three-dimensional (3D) nanomagnetism has attracted growing attention in recent years [1–5], driven not only by the fundamental interests in curved and topologically complex magnetic textures [6–8], but also their potential applications in next-generation spintronic devices [9–11]. Unlike conventional thin-film systems, where magnetization on planar surfaces can be easily probed, emerging 3D nanomagnetic architecture often involve tilted facets, patterned structures, or non-planar substrates. These structural complexities introduce additional challenges for magnetization imaging, demanding advanced detection techniques that combine height axial ($z$-axis) spatial resolution with sensitivity to arbitrary magnetization components.

Recent advances in electron- or X-ray-based magnetic tomography and laminography have demonstrated their capability in investigating 3D magnetization over complex geometries [12–18]. However, these approaches typically require large-scale facilities, highlighting the need for complementary, laboratory-accessible techniques that enable three-dimensional magnetization resolution in compact table-top setups. Magneto-optical Kerr effect (MOKE) microscopy has long served as a powerful and versatile tool for visualizing magnetic domains and dynamics, offering diffraction-limited resolution together with comparatively fast data acquisition and non-invasive nature [19]. While dark-field MOKE focused magnetometry has been utilized to characterize the magnetization vector in 3D magnetic nanostructures [20], direct MOKE microscopy for imaging 3D nanomagnetic samples faces intrinsic limitations. Importantly, the shallow depth of field (DOF) imposed by high-numerical-aperture objectives leads to blurred off-focus regions when imaging non-planar surfaces. Another challenge originates from the fact that both in-plane magnetization (longitudinal and transversal Kerr effects) and out-of-plane component (polar Kerr effect) contribute to the overall detected signal [21], making the interpretation of measurements more complicated on complex geometries.

Progress in computational optics and imaging reconstruction has opened promising pathways to address these challenges [22,23]. One simple but effective strategy can be through-focus scanning [24], which involves capturing a stack of images by varying the focus distance. Subsequently, the sharp and well-resolved regions within the focus range from each image are stitched together [25], effectively extending the depth of field and enabling high-quality imaging across non-planar surfaces. While proven practicable in various optical microscopy modalities, the applicability of through-focus scanning in MOKE microscopy has yet to be validated. Furthermore, as the effective incident angle of illumination varies with the local surface slope, the Kerr responses associated with in-plane and out-of-plane magnetization components need to be carefully evaluated.

In this work, we introduce an extended depth of field (EDOF) MOKE microscopy approach specifically designed for non-planar and tilted magnetic samples. By systematically integrating through-focus scanning with an image-stitching procedure, we reconstructed domain images that remain sharp and well-resolved across the entire field of view. Moreover, we analyzed the mechanisms that govern Kerr response under different illumination conditions, providing insights into how MOKE microscope can intrinsically probe in-plane magnetization along non-planar surfaces. This methodology not only extends the applicability of MOKE microscopy beyond flat films, but also establishes a versatile platform for studying complex magnetic systems with non-planar 3D geometries.

*Le Zhao: le.zhao@tuwien.ac.at
†Amalio Fernández-Pacheco: amalio.fernandez-pacheco@tuwien.ac.at

## II. EXPERIMENTAL METHODS

### A. MOKE microscopy setup

A home-built MOKE microscope was employed for the experiments. The illumination source was a high-power LED with four LED units, configured in a Köhler illumination scheme [26] to ensure uniform sample illumination. Two Glan-Taylor polarizers served as the polarizer and analyzer, respectively, and a $\lambda/4$ waveplate was used as the compensator. Imaging was performed using a 20×/0.5 objective and a digital CMOS camera. A piezo stage with $z$-axis movement was implemented for focusing and adjusting the distance between the sample surface and the objective, as illustrated in Fig. 1(a). A three-dimensional magnetic field was generated by a hexapole electromagnet, with its strength and orientation monitored using an *in-situ* 3D magnetic field sensor and three unidirectional Hall sensors mounted on the pole pieces of the electromagnet [27].

### B. Workflow for EDOF MOKE imaging

In our EDOF MOKE imaging scenario, a through-focus-scanning-based approach was adopted, with the overall workflow illustrated in Fig. 1(b) (see also section III.A-C for details). The central idea of through-focus scanning is to capture images at different focal planes, so that features located at various sample heights can be clearly resolved and stored in separate frames. A subsequent reconstruction step then stitches together the sharp features from each frame to generate an all-in-focus image. When adapted to MOKE imaging, however, this procedure requires specific modifications with respect to background subtraction. In conventional MOKE microscopy, a uniform background image taken at the saturated state of the sample is typically subtracted from the signal image to ensure a good signal-to-noise ratio. For non-planar structures, where magnetization information is distributed across different heights, a single background image is thus insufficient for all focal planes. Instead, a series of background images corresponding to different heights must be captured in advance. These images form a *Background Stack* that serves as a reference for image reconstruction. Once the external magnetic field is changed, another series of images within the range of interest is acquired, forming a *Signal Stack* that contains the non-uniform magnetization signals at different heights. After lateral drift compensation and height matching between those two stacks (see section III.A), background subtraction is performed between each background-signal image pair to obtain a *Subtracted Stack*. The sharpness of each image in the *Subtracted Stack* is then evaluated using a Sobel filter (which is generally a derivative of the image), and the final image can be reconstructed by stitching together the sharpest regions (see section III.B).

Practically, we tested the EDOF MOKE imaging method on a Ta(5)/Co$_{40}$Fe$_{40}$B$_{20}$(0.8)/MgO(1.1)/Ta(2) multilayer (all thicknesses are in nanometers) showing perpendicular magnetic anisotropy (PMA). The sample was tilted by 10° to mimic a non-planar structure. The range of through-focus scanning was $\pm 24\ \mu m$ with a step size of 200 nm. We also performed additional post-processing to the stitching results for further enhancing the signal-to-noise ratio (see section III.C). All those reconstruction products were then evaluated and compared quantitively in section III.D.

## III. ALGORITHM DESCRIPTION AND EVALUATION

### A. Drift compensation and height matching

Drifts in a MOKE microscope induced by the application of magnetic fields, mechanical instabilities or thermal fluctuations can compromise the image quality, especially after background subtraction. For a through-focus scanning scenario of three-dimensional structures, drifts along the $z$-direction may also lead to varying degrees of defocusing across the sample, which cannot be easily corrected only by two-dimensional image alignment. Therefore, matching each signal image to a specific background image taken at exactly the same height together with the lateral drift compensation must be achieved before image reconstruction. In order to enable automated processing, we sequentially applied a Sobel filter and a Gaussian filter to the signal image (Figs. 2(a-c)). Features used for alignment were then selected from a 21×21-pixel square region centered around each global maxima, as shown in Fig. 2(d). These steps help to guarantee that the selected region is always located within the in-focus area.

Alignment was performed between the selected signal image and a series of background images with the relative nominal height offset defined as $\Delta h$ (Fig. 2e). An algorithm based on the normalized cross-correlation coefficient was utilized using the Fiji software for drift compensation [28,29], with the searching window set to 5 pixels ($\approx 1.47\ \mu m$). Figs. 2(f) and 2(g) show the subtracted images before and after drift compensation, respectively, from which it can be observed that non-magnetic features cannot be eliminated through subtraction if the heights of the signal and background images are not identical.

*Le Zhao: le.zhao@tuwien.ac.at
†Amalio Fernández-Pacheco: amalio.fernandez-pacheco@tuwien.ac.at

Furthermore, the residual visibility of non-magnetic features can be utilized to identify the best-matching background image at a specific height. By applying a median filter which computes the local median of pixel values to the subtracted images and then a Gaussian filter to the resulting changes, the best-matching background image showing the least features can be effectively identified through a brightness analysis, as shown in Figs. 2(h-i). Fig. 2(j) shows the overall sample drift between *Signal Stack* and *Background Stack* during the imaging experiment. It is worth noting that drifts along lateral axes can also be compensated in real time at the hardware level by controlling the movement of the piezo stage along *x*- and *y*-directions.

## B. Heightmap-guided image stitching

### 1. Sharpness-based stitching

Following the background subtraction steps described above, a series of subtracted images were generated (Fig. 3(a)). Applying another Sobel filter allowed the identification of sharp regions in each subtracted image at different heights (Fig. 3(b)). By comparing the sharpness across regions, a heightmap was generated in which each pixel value corresponds to the height of the sharpest frame within the *Subtracted Stack* (Fig. 3(c)). Based on this heightmap, the sharpest regions from the *Subtracted Stack* were stitched together to reconstruct a fused domain image, as shown in Fig. 3(d).

The reconstructed image exhibits sharp domain edges across the entire field of view. However, since spatial information in magnetic systems is primarily concentrated at the domain edges rather than within the domains, directly comparing sharpness across the entire image may lead to difficulties in identifying the height indices for regions inside the magnetic domains, as indicated by the noise-like features within domain areas shown in Figs. 3(c-d). One practical way to address this issue consists of applying a Gaussian filter followed by a Sobel filter to Fig. 3(d) in order to identify the domain edges in the fused image. After subsequent binarization, an edge mask, as shown in Fig. 3(e), is generated. This mask allows featureless pixels in Fig. 3(c) to be filtered out, enabling a more reliable heightmap extraction, as discussed below.

### 2. Real-shape-guided stitching

In cases where the three-dimensional topography of the real sample is known in advance or qualitatively accessible, pixel-wise height information can be incorporated as an additional input, enabling heightmap fitting to further improve the reconstruction quality. Assuming that the experimentally measured heightmap and the intended profile are related through a linear transformation arising from imperfect calibration during device fabrication or inaccuracies in stage movement, this fitting procedure can then be achieved by minimizing the loss function

$$\mathcal{L}_1 \equiv \frac{1}{2}\sum_{\boldsymbol{p}\in\boldsymbol{P}}\left\{O(\boldsymbol{p}) - \left[\alpha I\big(\boldsymbol{R}(\theta)\times\beta\boldsymbol{T}(\boldsymbol{p})\big) + h_z\right]\right\}^2, (1)$$

where $\boldsymbol{p} = p(x, y)$ indicates each pixel of image within the entire field of view, $\boldsymbol{P}$ denotes the set of image pixels used for fitting, $O(\boldsymbol{p})$ and $I(\boldsymbol{p})$ represent the experimentally observed and a priori intended heightmap index at pixel $\boldsymbol{p}$, respectively. $\alpha$ is the $z$-direction distortion factor, $\boldsymbol{R}(\theta)$ is the rotation matrix, $\beta$ is the lateral expanding factor, $\boldsymbol{T}(\boldsymbol{p}) = \boldsymbol{p} - \boldsymbol{p_0}$ stands for the lateral translation operation where $\boldsymbol{p_0}$ is the rotation center, and $h_z$ is the overall offset along the $z$-direction.

The optimization process presented in Eq. (1) is schematically demonstrated in Figs. 3(f-i), where an artificial 3D surface serves as an example to illustrate the algorithm. Here the solid surfaces denote the (intermediate) fitting results, whereas the dotted surface corresponds to the surface derived directly from the experimentally observed heightmap. First, the lateral translation shown in Fig. 3(f) can be achieved by overlapping the representative reference points between the intended and the observed heightmaps. Next, as shown in Fig. 3(g), the lateral dimensions are calibrated using feature sizes. Subsequently, by performing an in-plane rotation around the $z$-axis, the intended and observed heightmaps are aligned laterally (Fig. 3(h)). This is then followed by a vertical translating and rescaling along the $z$-axis, after which the observed heightmap can be well fitted, as shown in Fig. 3(i).

In the case of our tilted test sample, however, the required transformations can be simplified and result in a direct 3D fitting of the tilted plane. After applying the edge mask (Fig. 3(e)) to the original heightmap (Fig. 3(c)), we obtained the masked heightmap shown in Fig. 3(j). The subsequent fitting process can be viewed as an extrapolation of the effective height indices onto the broader two-dimensional plane, as illustrated in Fig. 3(k). Based on the fitted heightmap, regions from subtracted images corresponding to different focal heights were stitched together. This produced a composite image with consistent clarity across the entire field of view, as shown in Fig. 3(l).

*Le Zhao: le.zhao@tuwien.ac.at

†Amalio Fernández-Pacheco: amalio.fernandez-pacheco@tuwien.ac.at

## C. PSF-based post-processing

We now turn to a more sophisticated approach beyond the sharpness-based method introduced above, aiming at further improving the reconstruction quality. From this perspective, the overall *Subtracted Stack* is treated as a 3D matrix. The fitting process in Eq. (1), followed by image stitching, thus essentially constitutes a resampling procedure as illustrated in Fig. 4(a). In the case of a tilted surface, the central resampling plane shown in green corresponds directly to the reconstructed image of Fig. 3(l). Furthermore, by manually adjusting the $z$-direction offset $h_z$ in Eq. (1), the *Resampled Stack* representing the stitched domain images with different degrees of blur is obtained, as shown in Fig. 4(b). Each blurred image can be thus considered as the convolution of the sharpest resampled frame ($h = 0$ μm) and a point spread function (PSF) that varies with the defocus distance $h$ due to the nature of the objective [30]. Owing to this, not only the sharpest frame but also the blurred frames of the *Resampled Stack* contain information about the magnetic states of the sample, which can be utilized collectively to improve imaging quality.

### 1. Simulated-PSF-based reconstruction

Using the built-in Richards & Wolf 3D optical model [31] in a Fiji plugin named "PSF Generator" [32,33], where vectorial-based diffraction is considered to evaluate the off-focus pattern, the PSF of our optical system was simulated with realistic optical parameters, as presented in Figs. 4(c-d). This generated PSF and the corresponding multi-level blurred *Resampled Stack* were utilized jointly for an iterative reconstruction performed by minimizing the loss function defined as

$$\mathcal{L}_2(\mathbf{R}_i) \equiv \frac{1}{N} \sum_{h \in H} \hat{\mathcal{L}}_2(\mathbf{R}_i, h) \equiv \frac{1}{N} \sum_{h \in H} \frac{1}{2} [\mathbf{E}_i(h)]^2$$
$$\equiv \frac{1}{N} \sum_{h \in H} \frac{1}{2} [\mathbf{R}_i * \mathbf{PSF}(h) - \mathbf{S}(h)]^2 \quad . \quad (2)$$

Here, $N$ is the total number of resampled frames at different heights used for optimization and $\hat{\mathcal{L}}_2(\mathbf{R}_i, h)$ is the loss function component corresponding to each specific defocus distance $h$. $\mathbf{R}_i$ is a two-dimensional matrix regarding the PSF-based reconstructed image at iteration $i$, "$*$" denotes the convolution operator, $\mathbf{PSF}(h)$ is the height-dependent point spread function with each pixel value given by $\mathrm{PSF}(h, x, y)$, and $\mathbf{S}(h)$ represents the image at defocus distance $h$ in the *Resampled Stack*. $\mathbf{E}_i(h) = \mathbf{R}_i * \mathbf{PSF}(h) - \mathbf{S}(h)$ thus defines the error function matrix between the forward-propagated blurred estimate and the resampled frame.

The height range $H$ limits the number of images used for optimization, and was set to ± 2 μm in this case, corresponding to a total of 21 resampled frames with the through-focus scanning step size of 200 nm.

For each iteration, the pixel-wise gradient of the loss function component $\hat{\mathcal{L}}_2(\mathbf{R}_i, h)$ with respect to a specific pixel $[\mathbf{R}_i]_{x,y} = R_i(x, y)$ can be expressed as:

$$\frac{\partial \hat{\mathcal{L}}_2(\mathbf{R}_i, h)}{\partial [\mathbf{R}_i]_{x,y}} = \sum_{p,q} [\mathbf{E}_i(h)]_{p,q} \cdot \frac{\partial [\mathbf{E}_i(h)]_{p,q}}{\partial R_i(x, y)} \quad . \quad (3)$$

Note that $[\mathbf{E}_i(h)]_{p,q}$ stands for the value of the error function matrix $\mathbf{E}_i(h)$ at pixel $(p, q)$. Coordinates $p$ and $q$ are introduced here to properly apply the chain rule, and are not necessarily equal to $x$ and $y$, respectively. Following the definition of convolution, the last term of Eq. (3) can be further expanded as

$$\frac{\partial [\mathbf{E}_i(h)]_{p,q}}{\partial R_i(x, y)} = \frac{\sum_{u,v} R_i(u, v) \cdot \mathrm{PSF}(h, p - u, q - v)}{\partial R_i(x, y)}$$
$$= \mathrm{PSF}(h, p - x, q - y)$$
$$= \widetilde{\mathrm{PSF}}(h, x - p, y - q) \quad . \quad (4)$$

Here, $\widetilde{\mathbf{PSF}}(h)$ denotes the PSF at height $h$ flipped along both the $x$- and $y$-directions, which can reduce to $\mathbf{PSF}(h)$ itself under symmetric coordinate selection. Substituting Eq. (4) into Eq. (3) and evaluating the derivative at the pixel level yields

$$\frac{\partial \hat{\mathcal{L}}_2(\mathbf{R}_i, h)}{\partial [\mathbf{R}_i]_{x,y}} = \sum_{p,q} [\mathbf{E}_i(h)]_{p,q} \cdot \widetilde{\mathrm{PSF}}(h, x - p, y - q)$$
$$= [\mathbf{E}_i(h) * \widetilde{\mathbf{PSF}}(h)]_{x,y} \quad , \quad (5)$$

which can be compactly written in matrix form as

$$\frac{\partial \hat{\mathcal{L}}_2(\mathbf{R}_i, h)}{\partial \mathbf{R}_i} = \mathbf{E}_i(h) * \widetilde{\mathbf{PSF}}(h) \quad . \quad (6)$$

Note that the flipped PSF here essentially reverses the forward convolution operation, thereby enabling efficient error backpropagation without resorting to explicit deconvolution procedures [34]. Taken together, the updated estimate $\mathbf{R}_{i+1}$ for the next iteration can be expressed as

$$\mathbf{R}_{i+1} = \mathbf{R}_i - \frac{\lambda}{N} \sum_{h \in H} \frac{\partial \hat{\mathcal{L}}_2(\mathbf{R}_i, h)}{\partial \mathbf{R}_i}$$
$$= \mathbf{R}_i - \frac{\lambda}{N} \sum_{h \in H} \mathbf{E}_i(h) * \widetilde{\mathbf{PSF}}(h) \quad , \quad (7)$$

*Le Zhao: le.zhao@tuwien.ac.at
†Amalio Fernández-Pacheco: amalio.fernandez-pacheco@tuwien.ac.at

where $\lambda$ is the nominal learning rate and is set to 1 throughout this work for simplicity.

After 100 iterations, the final reconstructed image shown in Fig. 4(e) was obtained, exhibiting improved image quality relative to the previous fitting result (Fig. 3(l)). To evaluate the robustness of this algorithm, we repeated the reconstruction after excluding the images within the intrinsic depth of field ($\pm 1$ μm) from the previous image stack ($\pm 2$ μm). The resulting reconstructed image is shown in Fig. 4(f).

### 2. Gaussian-PSF-based reconstruction

For comparison, a Gaussian-type PSF was estimated by evaluating the degree of blurring in the *Resampled Stack* (Fig. 4(b)) with respect to the sharpest resampled frame, as shown in Figs. 4(g-h). The resulting PSF exhibits qualitative agreement with the previously simulated one. Unlike the Simulated-PSF-based reconstruction, which relies on prior knowledge of the optical parameters of the imaging system, the present fitting-based approach employs a simplified Gaussian model to approximate the effective PSF directly from the *Resampled Stack* itself. As a result, this empirical approach is expected to account for integrated system imperfections such as illumination inhomogeneity and deviations in the effective numerical aperture, offering a practical alternative for image reconstruction. The corresponding reconstructed images based on the fitted Gaussian-type PSF are shown in Figs. 4(i-j), respectively.

### D. Comparison of different reconstruction methods

So far, we have presented the domain image reconstruction using several algorithms or post-processing methods including direct sharpness-based stitching, real-shape-guided fitting-based stitching, and the simulated- or fitted-PSF-based reconstruction. Beyond these methods, we further calculated the mean and median values of the resampled frames over a defocus distance range of $\pm 400$ nm as references described below. The performance of all the methods was evaluated by comparing the average signal-to-noise ratio (S/N) defined as the brightness difference between two opposite magnetization states over noise variation, together with the average edge sharpness $\bar{\eta}$ calculated at domain wall regions, as listed in Table 1. Based on these criteria, the reconstruction strategies can be summarized as follows:

[i] Sharpness-Based Stitching (see section III.B.1). As the most straight-forward method, direct stitching based on sharpness put together the sharpest segments from the *Subtracted Stack*, which can indeed ensure sharp edges after stitching. However, due to its limited capability to distinguish homogeneous in-domain regions, randomly distributed noise might be prominent and wrongly identified as target features. This can degrade the overall image quality, characterized by a small signal-to-noise ratio in the second column. Nevertheless, this reconstructed image and its corresponding heightmap serve as the foundation for all the other reconstruction processes presented here.

[ii] Real-Shape-Guided Stitching (see section III.B.2). With the help of prior height information from the real shape and the masked heightmap shown in Fig. 3(j), only the heights of reliable pixels are used for image reconstruction, while the heights of remaining pixels are extrapolated through fitting. This approach not only ensures a continuous and more realistic heightmap, but also yields a reconstructed image with enhanced quality including well-resolved in-domain regions. The smaller average edge sharpness $\bar{\eta}$ compared with the Sharpness-Based Stitching method can be attributed to noise and defects located next to domain edges during image acquisition.

[iii] Simulated-PSF-Based Reconstruction (see section III.C.1). Regarding the PSF-based reconstruction methods, since multiple frames from the *Resampled Stack* are collectively exploited, both the signal-to-noise ratio and average edge sharpness are improved compared with the Heightmap-Guided Stitching. Notably, when frames within the intrinsic depth of field are excluded for the reconstruction ($5^{th}$ column), the average signal-to-noise ratio increases further. This behavior can be attributed to the exclusion of incompletely compensated morphological defects and random noise, which manifest as high-frequency features and are more pronounced in sharper resampled frames. Meanwhile, the absence of the sharpest frames can also lead to a decrease in the average edge sharpness owing to the loss of high-frequency details.

[iv] Gaussian-PSF-Based Reconstruction (see section III.C.2). As the Gaussian model is rooted in geometric optics rather than wave optics, it cannot capture the diffraction-limited features of a realistic point spread function. This results in the suppression of high-frequency information, leading to smoother reconstructions with an increased signal-to-noise ratio but a reduced average edge sharpness compared with the previous method, where a physically accurate, diffraction-based PSF is employed.

*Le Zhao: le.zhao@tuwien.ac.at
†Amalio Fernández-Pacheco: amalio.fernandez-pacheco@tuwien.ac.at

Apart from the methods discussed above, two additional post-processing approaches, namely [v] Mean-Value Projection and [vi] Median-Value Projection, are included for reference. Five central frames selected from the *Resampled Stack* within the range of the intrinsic depth of field are considered, and the mean or median values at each pixel across these frames are computed to generate the final image. Using this approach, the averaging of resampled frames can further improve the signal-to-noise ratio by 28%. However, as the effective focus distance of these involved frames varies, this inevitably leads to a degradation of edge sharpness. In comparison, the median projection results in only a minor reduction in edge sharpness among all the methods evaluated , while still enhancing the signal-to-noise ratio to some extent by suppressing outlier pixels.

From this quantitative comparison, it can be concluded that for the sample under investigation, the Simulated-PSF-Based Reconstruction method provides the best edge sharpness among all the methods evaluated , while maintaining a decent signal-to-noise ratio. However, the PSF-based methods require dense vertical oversampling, with $z$ scanning steps much smaller than the intrinsic depth of field, as well as additional post-processing. By contrast, the Real-Shape-Guided Stitching yields reasonably good results with limited input data and modest computational cost, and is therefore primarily employed for subsequent sections of this work.

## IV. RESULTS AND DISCUSSION

### A. Labyrinthine domain pattern reconstruction

To validate the reliability and applicability of the EDOF MOKE imaging approach presented here, we conducted experiments on another sample composed of $Ta(5)/[Co_{40}Fe_{40}B_{20}(0.8)/MgO(1.1)/Ta(1)]_3/Ta(1)$ as a representative case. This sample was prepared near the spin reorientation transition at room temperature and exhibits a perpendicularly magnetized labyrinthine domain pattern in the demagnetized state, whose characteristic domain size is determined solely by its intrinsic magnetic parameters [35]. Such a property makes it suitable as a standard specimen for assessing imaging performance under various tilt angles.

Using the previously described stitching method (see section III.B.2), the magnetic domain morphologies at different sample tilt angles were reconstructed from acquired raw images, as shown in Fig. 5(a). It should be noted that the illumination cannot be maintained uniformly across the sample due to the tilted reflective surface, which leads to brightness inhomogeneities in these reconstructed images. To eliminate the influence of this effect on domain pattern characterization, the reconstructed images were then binarized for further analysis (Fig. 5(b)).

The domain pattern acquired from a tilted surface represents the projection of the actual domain structure along the tilted surface onto the horizontal plane. In other words, for the inherently isotropic labyrinthine domains, tilting the sample results in the compression of the pattern along the tilt direction by a factor of $\cos\theta$. This geometric effect can be quantitatively verified by comparing the domain periodicities along the $x$- and $y$-directions. Through performing a two-dimensional fast Fourier transform (2D FFT) on the binarized images, the intensity distributions in the frequency domain were obtained, as shown in Fig. 5(c). Radial profiles along $x$- and $y$-directions within a $\pm 15°$ range were extracted and averaged, yielding the directional intensity distributions as a function of spatial frequency, as illustrated in Figs. 5(d-e), respectively. Gaussian fitting was then applied to determine the peak positions, based on which the spatial frequencies at each tilt angle were obtained (see Fig. 5(f)). The horizontal spatial frequency $f_x$ increases with the tilt angle, whereas the vertical frequency $f_y$ remains mostly unchanged. The resulting frequency ratio $f_x/f_y$ was subsequently calculated and presented in Fig. 5(g), alongside the theoretical expectation of $1/\cos\theta$ . Within the experimental uncertainty, the measured frequency ratio agrees well with the theoretical prediction, further confirming the validity of the image-stitching procedure and demonstrating its applicability to magnetic imaging on non-planar surfaces.

### B. Kerr response to in-plane magnetization

In addition to the domain image reconstruction of PMA samples, the capability of imaging in-plane (IP) magnetization, as well as to investigate quantitatively the response to non-planar IP components, constitutes another important aspect for 3D magnetic imaging [36]. For this purpose, we employed a 40-nm-thick CoFeB film as a representative IP sample. In conventional MOKE microscopy, illumination and collection optical paths are usually perpendicular to the sample surface, and the detection of in-plane magnetization signals is made possible by selecting specific wave vectors of the incident light. However, for a tilted sample surface, the effective incident angle differs from that in the untilted configuration, giving rise to additional contributions to the detection sensitivity of in-plane magnetization [37,38]. The corresponding signal response and underlying mechanisms are discussed in detail below.

*Le Zhao: le.zhao@tuwien.ac.at

†Amalio Fernández-Pacheco: amalio.fernandez-pacheco@tuwien.ac.at

In our measurement system, the wave vector selection was implemented by inserting a rotatable aperture with one off-axis hole at the conjugate position of the objective's back focal plane. This configuration allows only the light passing through the hole to reach the sample surface, while blocking all the other components. The first column of Fig. 6(a) shows the images at the back focal plane of the objective during the measurement of a planar sample, after inserting an auxiliary Bertrand lens in the optical path before the tube lens and camera. Since those images are reflected once from the sample surface before being collected by the detector, they appear horizontally flipped. Here the image edges are defined by the size of objective's back pupil, as indicated by the yellow circles. As the sample's tilt angle increases, the LED images can be seen moving progressively to the right with respect to the front pupil indicated by the white circles, eventually moving beyond its boundary.

The corresponding hysteresis loops under each illumination and tilting condition are presented in Figs. 6(b-d), respectively. Note that the applied magnetic field was always aligned with the tilted sample plane by means of the hexapole electromagnet available in our setup [27]. Under the top-LED illumination, the absolute switching amplitude of the hysteresis loop exhibits a monotonic decrease with the increasing tilt angle, whereas under the bottom-LED illumination, it initially decreases slightly from 0° to 5° before dropping sharply to almost zero. Under all-LED illumination, the switching amplitude is nearly zero at the tilt angle of 0°, owing to the full compensation among the LED units. As the tilt angle increases from 0° to 5°, the equalization gets disrupted, resulting in the switching behavior similar to that observed under bottom-LED illumination. When the tilt angle increases further from 5° to 10°, the switching reverses its sign and begins to resemble that under top-LED illumination. These changes in switching amplitudes as a function of tilt angle are summarized in Fig. 6(e), where the $y$-axis represents the total switching amplitude under identical LED output. After dividing the switching amplitude of each hysteresis loop by the corresponding mean light intensity under the given illumination conditions, the normalized Kerr response for each measurement configuration was calculated, as presented in Fig. 6(f). Interestingly, although the absolute switching amplitude under all-LED illumination appears larger than that under top- and bottom-LED illumination at tilt angles of 5° and 10°, the corresponding Kerr response is in fact smaller. Furthermore, the Kerr response for all illumination conditions shifts toward the negative direction when the tilt angle increases from 0° to 5°. A detailed explanation of the underlying mechanisms is presented below.

To understand the trends described above, we schematically show in Figs. 7(a-d) the representative light paths for illumination during the measurement, incorporating the sample surface, objective lens, and the LED image at the back focal plane. Note that only geometric optics arguments are considered here. For a planar sample (Fig. 7(a)), the incident angles of light originating from the top-LED unit $\theta_T$ and bottom-LED unit $\theta_B$ are identical and approximately equal to arcsin(NA), with NA being the numerical aperture of the objective. As the tilt angle increases, the clockwise rotation of the sample surface leads to a decreasing $\theta_T' < \theta_T$ (Fig. 7(b)) and increasing $\theta_B' > \theta_B$ (Fig. 7(c)) for the top- and bottom-LED units, respectively. In other words, incident light from the top-LED unit becomes less effective in detecting in-plane magnetization, whereas that from the bottom-LED unit becomes more effective. This mechanism helps to explain the negative shift in the Kerr response observed at the tilt angle of 5° in Fig. 6(f).

Another important factor governing the observed hysteresis loops is the fraction of reflected light collected by the objective lens. As indicated in Fig. 6(a), tilting of the sample can cause the reflected image to fall partially outside the objective's front pupil, and thus it cannot be fully captured by the detector. Considering that the bottom-LED image is located on the left side of the figure and the reflected light propagates toward the right, this mechanism has a stronger impact on the bottom-LED illumination compared to the top-LED scheme. Under the top-LED illumination (Fig. 7(b)), as long as the tilt angle remains below $\theta_T \approx$ arcsin(NA), the reflected light is still largely collected by the objective, yet the variations in switching amplitude of hysteresis loops are primarily determined by the effective incident angle. Under the bottom-LED illumination (Fig. 7(c)), however, even a small tilt leads to a significant loss of the reflected light. This accounts for the observation that, although the normalized Kerr response increases with the tilt angle from 0° to 5° under bottom-LED illumination (Fig. 6(f)), the absolute switching amplitude still decreases (Fig. 6(e)), as there is an approximately 40% reduction in the illumination intensity. Moreover, for a sufficiently large tilt angle ($\theta_M$) determined by the numerical aperture and specific focus distance (Fig. 7(d)), none of the reflected light from the bottom-LED unit can be eventually collected by the objective. As a result, both the switching amplitude and the experimentally measured Kerr response approach zero at larger tilt angles of 15° and 20°.

*Le Zhao: le.zhao@tuwien.ac.at
†Amalio Fernández-Pacheco: amalio.fernandez-pacheco@tuwien.ac.at

Returning to the response under all-LED illumination, since the overall illumination changes moderately from 0° to 5°, the Kerr response at a tilt angle of 5° is primarily governed by the changes in the effective incident angle, which has the same sign as that of the bottom-LED illumination. For tilt angles larger than 10°, the absence of reflected light from the bottom-LED unit induces the dominance of top-LED unit, thereby reversing the overall measured switching direction. Owing to the competing contributions from the top- and bottom-LED illuminations, the Kerr response under all-LED illumination always lies between those obtained under the two individual illumination conditions, as revealed in Fig. 6(f). It is worth noting that although only the Köhler illumination path is depicted here, these arguments can be generalized to other configurations as the effective collecting angle of the objective remains unchanged.

The analysis above reveals the capability of detecting in-plane magnetized signals from a non-planar sample surface. Even under all-LED illumination without additional wave vectors selections, the MOKE microscope retains an intrinsic sensitivity to the in-plane magnetization along the tilting direction, as long as the tilting angle is smaller than arcsin(NA) defined by the objective. Building on this intrinsic sensitivity, we performed imaging of the in-plane magnetized domains under all-LED illumination, with the tilt angle set at 10° to maximize the magneto-optical Kerr response. The directly imaged domain pattern is shown in Fig. 7(e). Sharpness analysis (Fig. 7(f)) indicates that the domain boundaries are well defined near the center but blurred towards the edges. By employing the previously described through-focus scanning followed by image-stitching procedures (see section III.B.2), we obtained the reconstructed image shown in Fig. 7(g), which is well-resolved and clear across the entire field of view, with a 38% improvement in edge sharpness indicated in Fig. 7(h).

## V. CONCLUSIONS

In summary, we have presented an extended depth of field MOKE microscopy approach capable of capturing high-quality magnetic domain images on non-planar sample surfaces. By performing through-focus scanning followed by image-stitching based on sharpness analysis, we obtain magnetic domain images that are both sharp and well-resolved across the entire field of view. While the Real-Shape-Guided Stitching method proves effective in preserving edge sharpness and suppressing random noise, PSF-based reconstruction methods can further enhance both the signal-to-noise ratio and edge definition, particularly in regions affected by high-frequency noise or imaging defects. The reliability of such reconstruction strategy is quantitatively assessed by measuring and stitching domain images of a perpendicularly magnetized sample exhibiting isotropic periodic magnetic domains in the demagnetized state at different tilt angles. The observed spatial frequency of domains extracted from FFT analysis follows the expected $1/\cos\theta$ dependence on the tilt angle. The measurement response for an in-plane magnetized sample is also systematically investigated by varying the tilt angle under different illumination schemes. After analyzing the contributions of individual wave vectors from each LED unit, we have quantified how the effective incident angles influence the MOKE signal for in-plane magnetization, demonstrating quantitatively the intrinsic capability of MOKE microscopy to resolve in-plane magnetic signals on non-planar surfaces.

Although demonstrated here as proof-of-concept cases with intentionally tilted flat samples, the presented methods and findings can be readily extended to more complicated 3D geometries. Compared to X-ray-based magnetic domain reconstruction techniques, the simplicity and flexibility of this laboratory-accessible method allow it to be easily implemented in standard MOKE microscopy setups with only moderate modifications, extending its availability to a wide research community. Due to its relatively fast imaging speeds, it can also facilitate real-time observations of magnetic domain evolution and dynamics.

Additionally, it is worth noting that light reflected from off-focus surface regions may interfere with the reconstruction process. An effective solution in that case, inspired by the widely used confocal microscopy technique in biological research [39], could comprise inserting a pinhole together with auxiliary lenses into the imaging path before the camera [40]. This configuration selectively collects signals originating only from the focal plane, which are subsequently used for image reconstruction. There are also additional algorithms for reconstruction involving neural networks and deep learning [41], but these are beyond the scope of the present manuscript. Future improvements could also involve synchronization with pulsed magnetic fields or the implementation of real-time image processing, to enable the time-resolved measurement of magnetization dynamics. We believe that the continued development of such integrated imaging strategies promises to provide a powerful tool for investigating surface magnetization states in 3D geometries, thereby opening new avenues for tabletop studies of non-planar magnetic nanostructures and spintronic devices.

*Le Zhao: le.zhao@tuwien.ac.at
†Amalio Fernández-Pacheco: amalio.fernandez-pacheco@tuwien.ac.at

## ACKNOWLEDGMENTS

This work was supported by the European Community under the Horizon 2020 Program, Contract No. 101001290 (3DNANOMAG). This research was funded in whole or in part by the Austrian Science Fund (FWF) [10.55776/PIN1629824], through the TOMOCHIRAL project. Funding from TU Wien's Blue Sky "Phoenix" program is acknowledged. We also thank Dr. Jakub Mateusz Jurczyk for his support during the measurements.

*Le Zhao: le.zhao@tuwien.ac.at

†Amalio Fernández-Pacheco: amalio.fernandez-pacheco@tuwien.ac.at

*Le Zhao: le.zhao@tuwien.ac.at

†Amalio Fernández-Pacheco: amalio.fernandez-pacheco@tuwien.ac.at

**(a)**

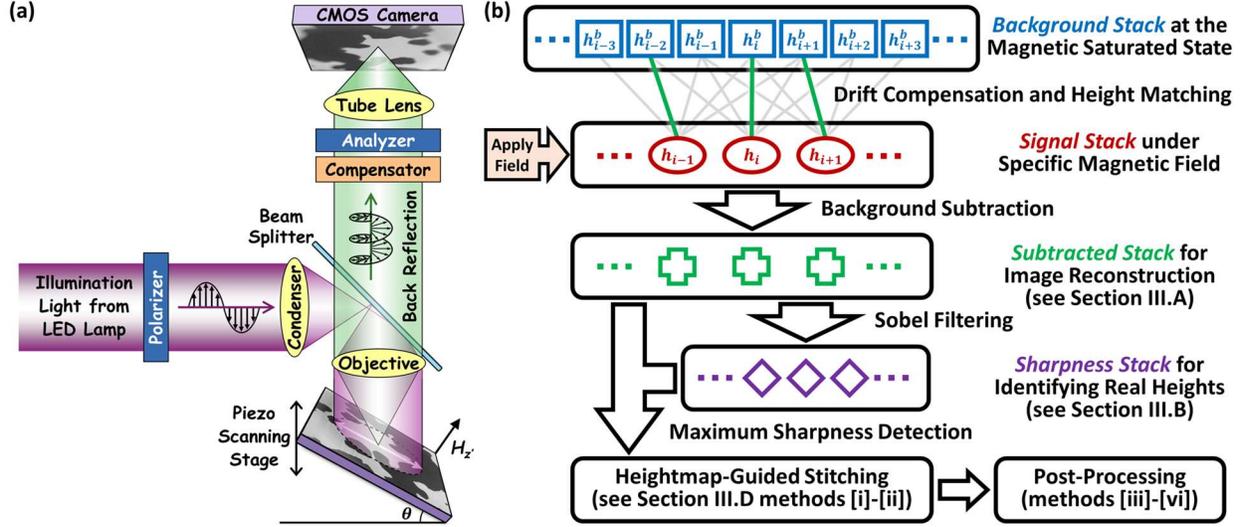

FIG. 1. Experimental setup and workflow for extended depth of field (EDOF) magneto-optical Kerr effect (MOKE) microscopy. (a) Optical path of the wide-field MOKE microscopy system. The flat magnetic sample was mounted on a tilted sample holder to imitate a non-planar surface. (b) Flowcharts illustrating the EDOF MOKE imaging process. Unlike conventional MOKE microscopy, where the sample position remains fixed while the magnetic field is varied, here a series of images at different heights $h_i$ are acquired by moving the sample along $z$-axis for each magnetic field value. This procedure ensures that magnetization variations across the non-planar surface are fully resolved. The resulting *Signal Stack* is then compared with a pre-acquired *Background Stack* that captures the sample features at different heights $h_i^b$. After drift compensation, height matching and background subtraction, the *Subtracted Stack* used for image reconstruction is prepared (see section III.A). Finally, Sobel-filter-based sharpness detection is used to guide the reconstruction of an image that is uniformly sharp and well resolved across the entire field of view (see section III.B). Additional post-processing is applied to further improve the image quality (see sections III.C-D).

*Le Zhao: le.zhao@tuwien.ac.at

†Amalio Fernández-Pacheco: amalio.fernandez-pacheco@tuwien.ac.at

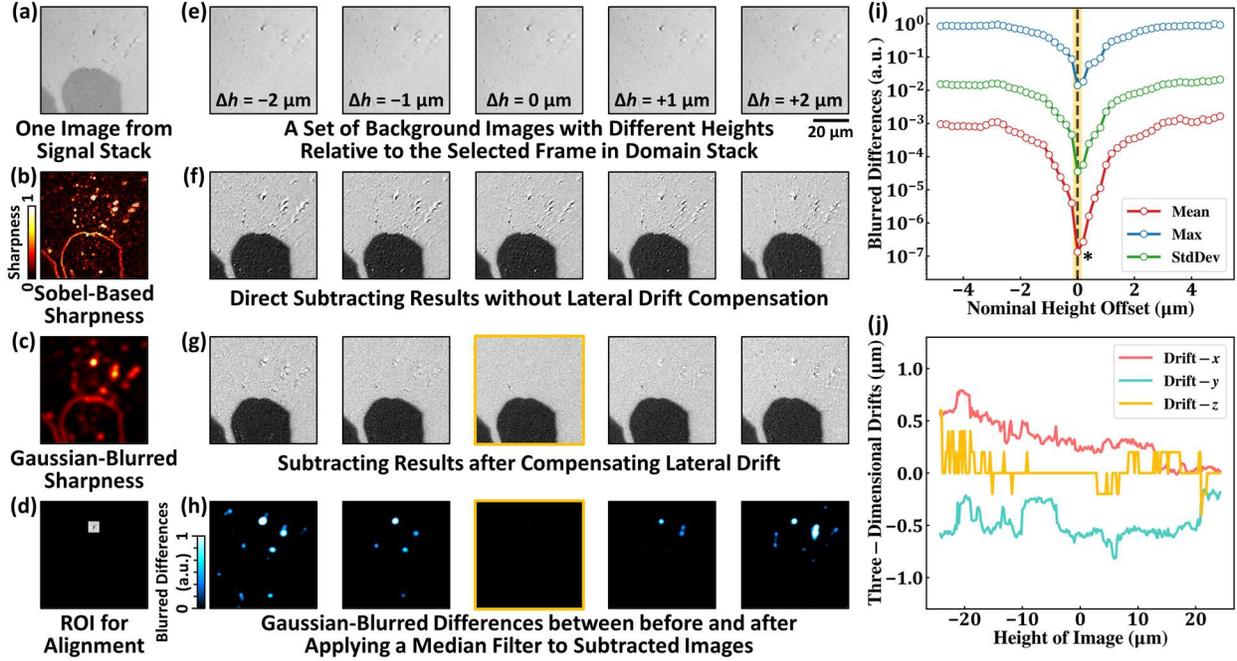

FIG. 2. Drift compensation and height matching procedures, illustrated using a Ta/CoFeB/MgO/Ta sample exhibiting perpendicular magnetic anisotropy. (a-d) Identification of region of interest (ROI) for alignment. (a) Selected signal image as an example. (b) Application of a Sobel filter allows to identify structural features. (c) Subsequent application of a Gaussian filter removes random noise. (d) Selected feature region for alignment. (e-i) Matching the best background image to the signal image. (e) Background images acquired at different heights relative to the original image height ($\Delta h = h_i - h_j^b$). (f) Background-corrected images without drift compensation. Extra features in the images appear due to small lateral drifts along $x$- and $y$-directions, together with the differences of focus distances between the signal image and background images. (g) Subtracted images after drift compensation. Though having the lateral drifts compensated, changes in imaging heights between the signal and background images still result in minor topography-related features. (h) Gaussian-filtered results after calculating the changes induced by a median filter, among which the optimal background image with minimum extra features is picked and marked by the yellow frame. (i) Brightness analysis results for a series of images prepared as in (h), with the optimal height offsets indicated by the minimum of the blurred differences. (j) Computed 3D drifts between the entire *Signal Stack* and *Background Stack* during the experiment. Figures (a-h) share the same scale bar, and Figures (b-c) share the same colormap.

*Le Zhao: le.zhao@tuwien.ac.at

†Amalio Fernández-Pacheco: amalio.fernandez-pacheco@tuwien.ac.at

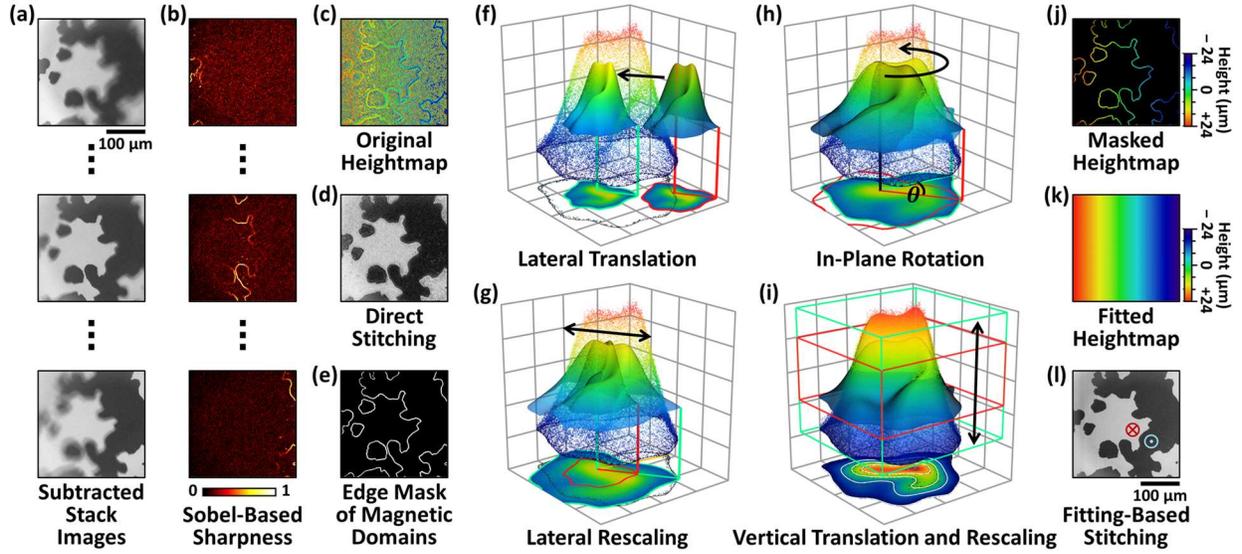

FIG. 3. Heightmap-guided image stitching methods. (a) Representative images from the *Subtracted Stack* at different heights $h_i$. (b) Sharpness evaluation of each subtracted image using a Sobel filter. (c) Heightmap obtained by assigning, for each region, the height index of the sharpest image within the *Subtracted Stack*. (d) Domain image reconstructed by stitching together the sharpest regions according to the heightmap in (c). (e) Edge mask generated from (d). (f-i) General optimization sequence for heightmap fitting. (f) Lateral translation is achieved by aligning the observed and intended reference points, corresponding to the $T(p)$ operation in Eq. (1). (g) Lateral rescaling by calibrating featured lateral sizes, corresponding to the expanding factor $\beta$ in Eq. (1). (h) In-plane rotation around $z$-axis, corresponding to the rotation matrix $R(\theta)$ in Eq. (1). (i) Vertical translation and rescaling along $z$-axis, corresponding to the factor $h_z$ and $\alpha$ in Eq. (1), respectively. (j) Masked heightmap obtained by applying the edge mask in (e) to the original heightmap in (c). (k) Fitted two-dimensional heightmap. (l) Final composite image with enhanced clarity. Figures (a-e) and (j-l) share the same scale bar, and Figures (c), (j) and (k) share the same colormap.

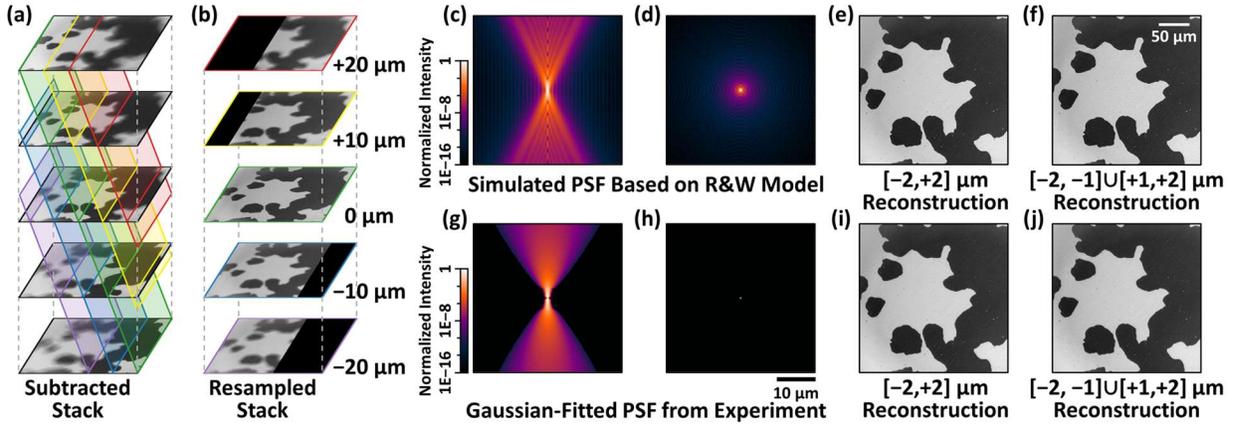

FIG. 4. PSF-based image reconstruction methods. (a) Fitting-based stitching can be regarded as the resampling of the *Subtracted Stack*, in which the intersection lines between the colored fitting plane and the subtracted frames are extracted to generate the resampled frames. (b) Resampled frames exhibiting various degrees of blur. (c) Side view ($y = 0$) and (d) top view ($z = 0$) of the point spread function (PSF) simulated using the Richards & Wolf 3D optical model. (e-f) Reconstructed images obtained through iterative optimization with the simulated PSF and resampled frames within a range of (e) $\pm 2$ μm; and (f) $\pm 2$ μm, but with the sharpest frames excluded. (g) Side view ($y = 0$) and (h) top view ($z = 0$) of the PSF derived by fitting the resampled frames. (i-j) Reconstructed images obtained through iterative optimization with the Gaussian-fitted PSF and resampled frames within a range of (i) $\pm 2$ μm; and (j) $\pm 2$ μm, but with the sharpest frames excluded. Figures (c), (d), (g) and (h) share the same scale bar together with the colormap, and Figures (e), (f), (i) and (j) share the same scale bar.

*Le Zhao: le.zhao@tuwien.ac.at

†Amalio Fernández-Pacheco: amalio.fernandez-pacheco@tuwien.ac.at

Table 1. Performance comparison of different reconstruction methods introduced in the main text. Rows 4 and 5 list the average intensity and standard deviation measured in the bright and dark domain regions, respectively, together with the corresponding noise ratio in percentage. Row 6 summarizes the average signal-to-noise ratio (S/N), serving as one quantitative metric for reconstruction quality. Here, the signal is defined as the average brightness difference between the bright and dark domain regions, while the noise is systematically evaluated over the entire image based on the deviations. The last two rows present the absolute Laplacian-filtered domain image, highlighting the domain boundaries, as well as the corresponding average edge sharpness $\bar{\eta}$ calculated by averaging pixel values within the edge mask shown in Fig. 3(e). For both quality indicators, S/N and $\bar{\eta}$, higher values indicate better performance and are highlighted using a red-yellow-green color scale. As inspired by method [v], averaging only five images yields a high S/N, but at the expense of substantially reduced edge sharpness. In contrast, method [iii] achieves the highest edge sharpness using 21 input images while maintaining a reasonable S/N, demonstrating the effectiveness of the PSF-based reconstruction approach.

| Algorithms and Post-Processing Methods | [i] Sharpness-Based Stitching | [ii] Real-Shape-Guided Stitching | [iii] Simulated-PSF-Based Reconstruction | | [iv] Gaussian-PSF-Based Reconstruction | | [v] Mean-Value Projection | [vi] Median-Value Projection |
|---|---|---|---|---|---|---|---|---|
| Range of Used Resampled Frames ($h$) | | 0 µm (One Frame) | [−2,+2] µm (21 Frames) | [−2,−1] ∪ [+1,+2] µm (12 Frames) | [-2,+2] µm (21 Frames) | [−2,−1] ∪ [+1,+2] µm (12 Frames) | [−0.4,+0.4] µm (5 Frames) | |
| Reconstructed Domain Image | 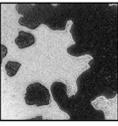 | 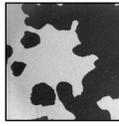 | 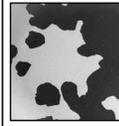 | 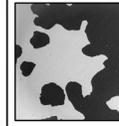 | 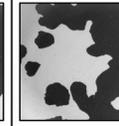 | 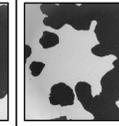 | 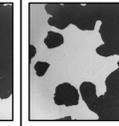 | 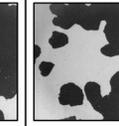 |
| Bright-region Intensity | 43418 ± 7239 (±16.7%) | 43372 ± 4577 (±10.6%) | 44021 ± 4025 (±9.1%) | 43934 ± 3913 (±8.9%) | 44050 ± 3891 (±8.8%) | 43958 ± 3777 (±8.6%) | 43510 ± 3693 (±8.5%) | 43399 ± 3791 (±8.7%) |
| Dark-region Intensity | 12311 ± 7524 (±61.1%) | 14236 ± 4031 (±28.3%) | 14608 ± 3432 (±23.5%) | 14679 ± 3296 (±22.5%) | 14623 ± 3246 (±22.2%) | 14667 ± 3100 (±21.1%) | 14615 ± 2973 (±20.3%) | 14563 ± 3087 (±21.2%) |
| Average Signal-to-Noise Ratio (S/N) | 4.208 | 6.783 | 7.902 | 8.129 | 8.259 | 8.529 | 8.677 | 8.395 |
| Absolute Laplacian-Filtered Domain Image | 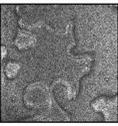 | 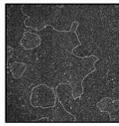 | 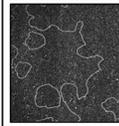 | 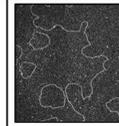 | 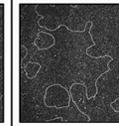 | 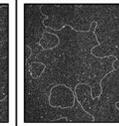 | 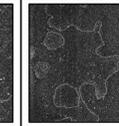 | 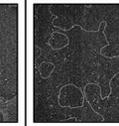 |
| Average Edge Sharpness ($\bar{\eta}$) | 0.743 | 0.673 | 0.787 | 0.767 | 0.772 | 0.745 | 0.630 | 0.663 |

*Le Zhao: le.zhao@tuwien.ac.at

†Amalio Fernández-Pacheco: amalio.fernandez-pacheco@tuwien.ac.at

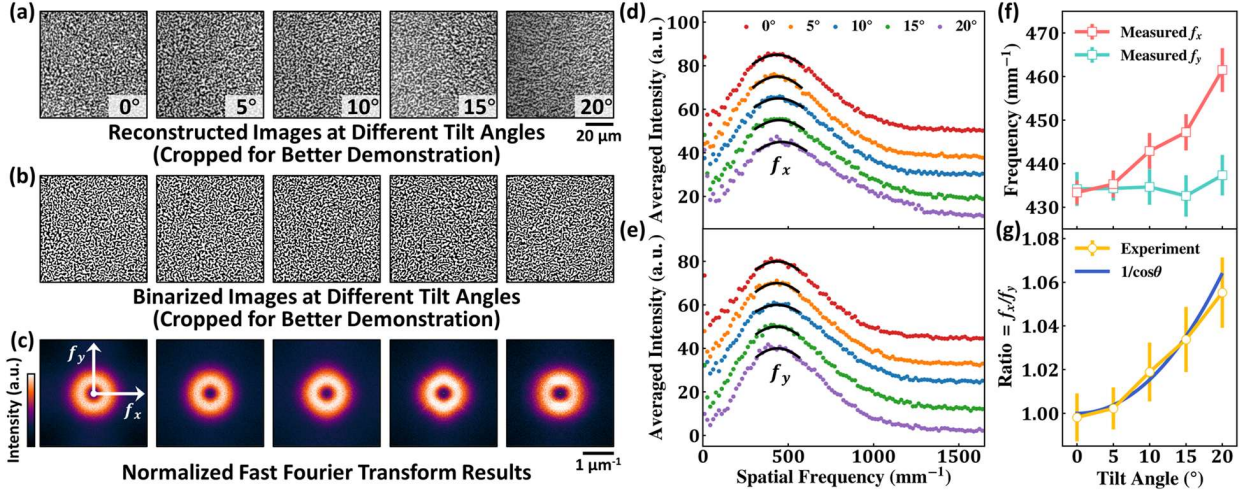

FIG. 5. Out-of-plane labyrinthine domain pattern reconstruction, demonstrated using a magnetic Ta/[CoFeB/MgO/Ta]$_3$ multilayer with PMA and close to the spin reorientation transition. (a) Reconstructed labyrinthine domain images acquired at different tilt angles. (b) Binarization results of (a). (c) Two-dimensional Fast Fourier Transform (2D FFT) results of the binarized domain images. (d) Horizontal and (e) vertical profiles of the 2D FFT results. To reduce fitting errors, pixels within ± 15° regions along either the *x*- or *y*-direction were averaged. (f) Summary of the spatial frequencies of labyrinthine domains along the *x*- and *y*-directions as a function of tilt angle. (g) Spatial frequency ratio $f_x/f_y$ as a function of tilt angle, together with the theoretical $1/\cos\theta$ relation based on a projected geometry. Figures (a) and (b) share the same scale bar.

*Le Zhao: le.zhao@tuwien.ac.at

†Amalio Fernández-Pacheco: amalio.fernandez-pacheco@tuwien.ac.at

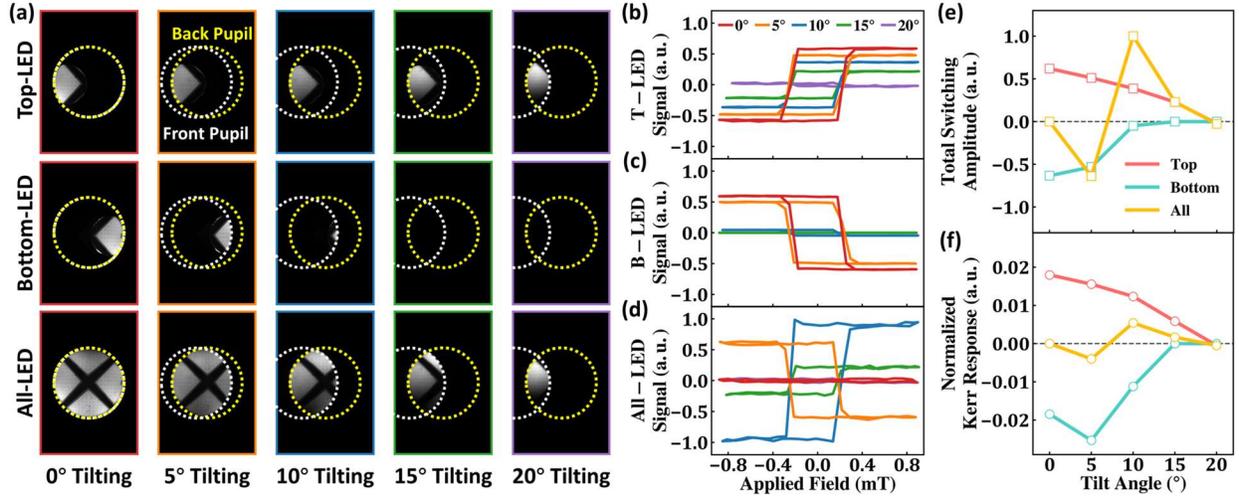

FIG. 6. Measurement results of a tilted in-plane magnetized film at different tilt angles, demonstrated using a thick CoFeB layer exhibiting in-plane magnetic anisotropy. (a) Conoscopic images at the back focal plane of the objective lens captured with different tilt angles of the sample after the insertion of a Bertrand lens. Here, yellow and white dashed circles indicate the edge of the back and front pupils of objective, respectively. (b-d) Measured hysteresis loops at different tilt angles using (b) top-, (c) bottom- and (d) all-LED units. (e) Summary of the total switching amplitudes at different tilt angles obtained from the corresponding hysteresis loops. (f) Calculated normalized Kerr response at different tilt angles.

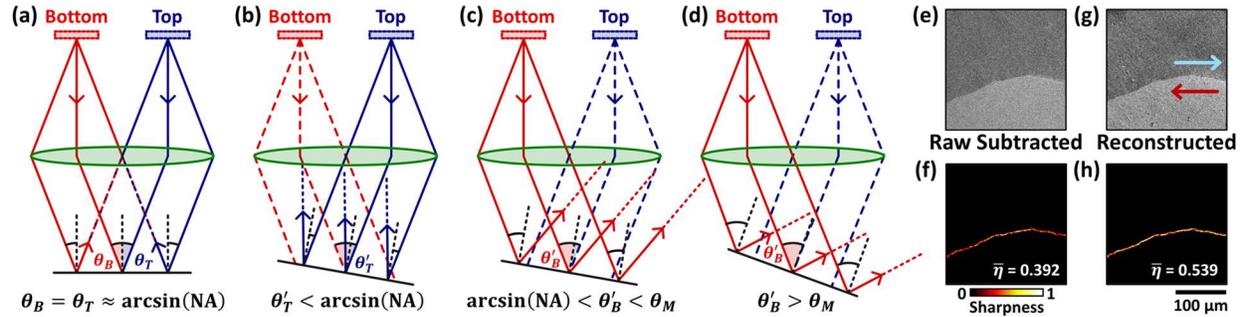

FIG. 7. Light-path sketches and in-plane domain image reconstruction. (a) For a planar sample surface, the effective incident angles for the top- and bottom-LED units are identical. (b-c) When the sample surface is tilted clockwise, the effective incident angle for top-LED unit decreases (b), while that for bottom-LED unit increases (c). (d) If the tilt angle becomes sufficiently large, no reflected light under bottom-LED illumination can be eventually collected by the objective due to its finite numerical aperture. (e) Directly imaged in-plane magnetic domains on a tilted film under all-LED illumination. (f) Edge sharpness of image (e). (g) Reconstructed in-plane domain image with clear details across the entire field of view. (h) Edge sharpness of image (g). Figures (e-h) share the same scale bar, and Figures (f) and (h) share the same colormap.

*Le Zhao: le.zhao@tuwien.ac.at

†Amalio Fernández-Pacheco: amalio.fernandez-pacheco@tuwien.ac.at